\documentclass[superscriptaddress,pra,twocolumn]{revtex4}
\usepackage{epsfig}
\usepackage{delarray}
\usepackage{amsmath, amssymb}
\usepackage{bm}
\usepackage{dsfont}

\begin{document}
\title{Quantum Pseudo-Telepathy in Spin Systems: Magic Square Game Under Magnetic Fields and Dzyaloshinskii-Moriya Interaction}

\author{Fatih Ozaydin}
\email{MansurSah@gmail.com} 
\affiliation{Department of Information Technologies, Isik University, Sile, Istanbul, 34980, Turkey}

\date{\today}

\begin{abstract}
Dzyaloshinskii-Moriya (DM) interaction has been proven to excite entanglement of spin systems, enhancing the capability of realizing various quantum tasks such as teleportation.
In this work,  we consider the DM interaction -to the best of our knowledge, for the first time in quantum game theory.
We study the winning probability of magic square game played with the thermal entangled state of spin models under external magnetic fields and DM interaction.
We analytically show that although DM interaction excites the entanglement of the system as expected, it surprizingly reduces the winning probability of the game, acting like temperature or magnetic fields, and also show that the effects of DM interaction and inhomogeneous magnetic field on the winning probability are identical.
In addition, we show that XXZ model is considerably more robust than XX model against destructive effects.
Our results can open new insights for quantum information processing with spin systems.\\
\textit{\textbf{Keywords:} magic square game, quantum game theory; Dzyaloshinskii-Moriya interaction}
\end{abstract}

\maketitle

\section{Introduction}
Game theory has recently attracted an intense attention in quantum information science due to its importance in constructing scenarios where quantum mechanical resources, communication and operations outperform their classical counterparts.
Meyer showed that a quantum player \textit{(i.e. a player who can implement a quantum strategy utilizing quantum superposition)} can beat a classical player with certainty in a coin tossing game \cite{Meyer99PRL} and introducing quantization to the famous Prisoners' Dilemma game, Eisert et al. showed that the presence of entanglement gives rise to superior performance \cite{Eisert99PRL}.
Following these pioneering works, a huge effort has been devoted to bring the role of non-classical resources out in outperforming the classical resources in game theory via considering various games \cite{Benjamin2001PRL,Chen2002QIP,Lee2003PLA,Lee2003PRA,Piotrowski2003PhysA,Ozdemir2004PLA325, Shimamura2004PLA,Ozdemir2004PLA333,Nawaz2004JPA,Paternostro2005NJP,Ozdemir2007NJP,Piotrowski2008PhysA,Brunner2013NComm,Kerenidis2015PRL,Scarani2015PRA}.
Such an interesting game is the so-called magic square game (MSG), which we detail in the next section.
MSG is played by two players against a referee and utilizing a pre-shared quantum system of 4 qubits in a specific pure entangled state, players can win the game with unit probability, while players equipped only with classical resources can achieve a winning probability at most $8/9$ \cite{Aravind2002FoP,Brassard2005FoP}.
In practical applications, however, due to inevitable interactions of the state with the environment \cite{Gawron2008IJQI,Ramzan2010QIP,Fialik2012QIP}, or even accelerating players \cite{Gawron2016ActaB}, the entanglement of the state may decrease considerably, leading to the decrease of the winning probability even far below the classical limit.
On the other hand, Pawela et al. have recently showed that in spite of decoherence, the winning probability can be enhanced if each party performs local operations (determined by semi-definite programming) to their qubits \cite{Pawela2013Plos1}.
There is also an increasing interest on playing games with Heisenberg spin chains  \cite{Jarek2012QIP} and thermal entanglement \cite{Dajka2015Plos1}. \\
\indent In spin systems, the presence of anisotropic antisymmetric exchange between neighboring magnetic spins -the so called \textit{Dzyaloshinskii-Moriya interaction} \cite{Dzyaloshinskii,Moriya,Tatara2016PRL} plays an important role in the entanglement dynamics.
Zhang showed that Dzyaloshinskii-Moriya (DM)interaction excites the entanglement of a two qubit Heisenberg chain, and therefore it enhances the capability of realizing quantum tasks requiring entanglement such as teleportation \cite{Zhang2007PRA}.
Following this work, DM interaction has been studied intensely from the perspective of entanglement dynamics of various spin systems, showing that the presence of DM interaction overwhelms the destructive effects of temperature and magnetic fields by exciting the entanglement, i.e. increasing the amount of entanglement measures such as negativity and concurrence \cite{DMRefs1,DMRefs2,DMRefs3,DMRefs4,DMRefs5,DMRefs6,DMRefs7,DMRefs8}.
Very recently, the role of DM interaction in quantum metrology has also been studied, showing that DM interaction enhances the precision of parameter estimation \cite{MetroRef1,MetroRef2,MetroRef3}.
To the best of our knowledge, however, DM interaction has not been considered in game theory. \\
\indent In this paper, considering a thermalized spin system with DM interaction under homogeneous and inhomogeneous external magnetic fields, we study the influence of DM interaction and magnetic fields on the winning probability of the magic square game.
We show that although DM interaction excites the amount of entanglement of the system, surprizingly it does not increase but decreases the winning probability.
We also report that although resulting in different density matrices and entanglement values, inhomogeneous magnetic field and DM interaction of the same strength yield the same winning probabilites.
This paper is organized as follows. In Section II, we briefly introduce the magic square game.
In Section III, we present our model to play the game and in Section IV we present our results and discussions.
Finally we conclude in Section V.

\begin{figure}[t!]
  \centering
      \includegraphics[width=0.45\textwidth]{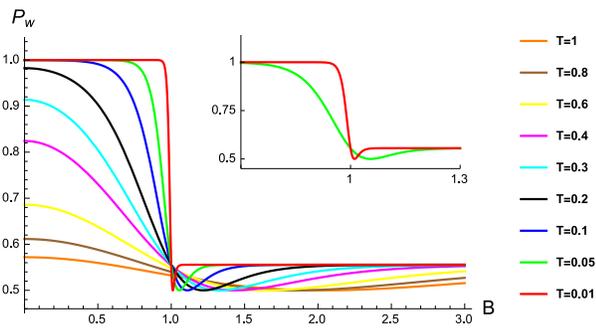}
  \caption{\label{fig:B}(Color online). Probability of winning the magic square game, $P_w$ with respect to external homogeneous magnetic field, $B$ at various temperatures from $T=0.01$ to $T=1$, in the units of Boltzmann constant $k$. Inset is focused on the region around $B=1$ for a more clear observation on the sudden change point. $P_w=8/9$ is the classical limit, therefore for low temperatures, quantum resources outperform the classical resources even under external magnetic field $B\leq 1$.}
\end{figure}

\section{Magic Square Game}
The game is played on a 3 by 3 matrix of binary entries by Alice and Bob against a referee, who gives the number of the row to Alice and the number of the column to Bob.
In order to win the game, the sum of the entries of the row given by Alice must be even, the sum the entries of the column given Bob must be odd and the intersection bit must agree.
Before starting the game, Alice and Bob can agree on any strategy and can share any physical resources but once the game starts, that is they are spatially separated and the referee gives them the numbers of the row and the column, they cannot communicate or share new resources.
Having shared classical resources, the highest winning probability via the best strategy they can achieve is $8/9$.
However, sharing quantum mechanical resources, in particular a specific entangled state of four qubits, they can achieve a unit winning probability.
The four-qubit state to share, first and second by Alice, and third and fourth by Bob is
\begin{equation}
|\Psi\rangle = {1 \over 2} ( |0011\rangle + |1100\rangle - |0110\rangle - |1001\rangle ).
\end{equation}

\noindent When the game starts and referee gives the number of the row, $m$ to Alice and the number of the column, $n$ to Bob, each of them applies one of the unitary operators $A_m$ and $B_n$ below to the two qubits they possess \\

\noindent $A_1 ={1 \over \sqrt{2}}
\left[
  \begin{array}{cccc}
    i &  \ \ 0 & 0 & 1 \\
    0 & -i & 1 & 0 \\
    0 &  \ \ i & 1 & 0 \\
    1 &  \ \ 0 & 0 & i \\
  \end{array}
\right], \ \
A_2 = {1 \over 2} \left[
  \begin{array}{cccc}
   \ \  i & 1 & \ \  1 & \ \ i \\
       -i & 1 &     -1 & \ \ i \\
   \ \  i & 1 &     -1 & -i \\
       -i & 1 & \ \  1 & -i \\
  \end{array}
\right],\\
A_3 = {1 \over 2} \left[
  \begin{array}{cccc}
       -1 &    -1 &     -1 & \ \ 1 \\
   \ \  1 & \ \ 1 &     -1 & \ \ 1 \\
   \ \  1 &    -1 &  \ \ 1 & \ \ 1 \\
   \ \  1 &    -1 &     -1 &    -1 \\
  \end{array}
\right], \ \ \mbox{and} \\
B_1 = {1 \over 2}
\left[
  \begin{array}{cccc}
    \ \  i &    -i & \ \ 1 & \ \ 1 \\
        -1 &    -i & \ \ 1 &    -1 \\
    \ \  1 & \ \ 1 &    -i & \ \ i \\
        -i & \ \ i & \ \ 1 & \ \ 1 \\
  \end{array}
\right], \
B_2 = {1 \over 2} \left[
  \begin{array}{cccc}
      -1 & \ \  i & 1 & \ \ i \\
  \ \  1 & \ \  i & 1 &    -i \\
   \ \ 1 &     -i & 1 & \ \ i \\
      -1 &     -i & 1 &    -i \\
  \end{array}
\right], \\
B_3 = {1 \over \sqrt{2}} \left[
  \begin{array}{cccc}
  \ \  1 & 0 & \ \ 0 & 1 \\
      -1 & 0 & \ \ 0 & 1 \\
  \ \  0 & 1 & \ \ 1 & 0 \\
  \ \  0 & 1 &    -1 & 0 \\
  \end{array}
\right],
$\\

\begin{figure}[t!]
  \centering
      \includegraphics[width=0.45\textwidth]{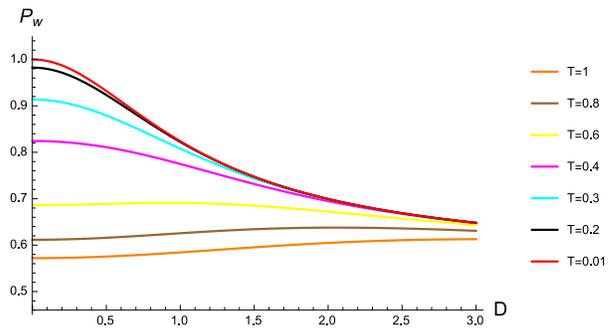}
  \caption{\label{fig:D}(Color online). Probability of winning the magic square game, $P_w$ with respect to Dzyaloshinskii-Moriya (DM) interaction, $D$ at various temperatures from $T=0.01$ to $T=1$, in the units of Boltzmann constant $k$. DM interaction counterintuitively does not increase but decreases the performance of the quantum task. However, since $P_w=8/9$ is the classical limit, for low temperatures, quantum resources outperform the classical resources even with a DM interaction of strength $D \leq 0.75$. Since the effects of $D$ and external magnetic field, $b$ on $P_w$ are identical, the same plot applies to $b$ as well.}
\end{figure}

\noindent $m$ and $n$ running from 1 to 3. After applying the corresponding operations, each determines the two bits and the third bits are found according to the parity conditions.
Note that the four-qubit state $|\Psi\rangle$ is actually the product state of two Bell pairs i.e. ${1 \over 2}[(|01\rangle - |10\rangle)\otimes (|01\rangle - |10\rangle)]$, such that the first and third qubits belong to Alice, and the second and fourth qubits belong to Bob.

For a given arbitrary system described by the density matrix $\sigma$, the winning probabilities of each row $m$ and column $n$ can be calculated as $P_{m,n}(\sigma) = tr(\sigma_f \Sigma_{m, n}|\phi_{m,n}\rangle \langle\phi_{m,n}|)$, where $\sigma_f = (A_m \otimes B_n) \sigma (A^{\dagger}_m \otimes B^{\dagger}_n)$, $\phi_{m,n}$ are the states implying success, and the winning probability of the given system can be obtained by averaging the winning probabilities over rows and columns, i.e. $P_w(\sigma) = {1 \over 9} \Sigma_{m,n} P_{m,n}$ \cite{Gawron2008IJQI}.

\section{Model}
To construct our model, we consider two antiferromagnetic XXZ spin chains of two qubits each, in particular with the coupling constants $J_x=J_y=1$ and $J_z \geq 0$ where the spin model reduces to XX model for $J_z =0$.
The Hamiltonian we consider ofour model is
$ {1 \over 2} [\sigma_{x}^{1}\sigma_{x}^{2} + \sigma_{y}^{1}\sigma_{y}^{2} + J_z \sigma_{z}^{1}\sigma_{z}^{2} + \overrightarrow{D} \cdot (\overrightarrow{\sigma}^{i} \times \overrightarrow{\sigma}^{i+1})   + (B+b)\sigma_{z}^{1}    +  (B-b)\sigma_{z}^{2}  ],$ where $B$ and $b$ are the strengths of the homogeneous and inhomogeneous magnetic fields, respectively and $D$ is the strength of the DM interaction which we will choose in $z$ direction for simplicity, resulting in the Hamiltonian

\begin{multline}
H = {1 \over 2} [\sigma_{x}^{1}\sigma_{x}^{2} + \sigma_{y}^{1}\sigma_{y}^{2} + J_z \sigma_{z}^{1}\sigma_{z}^{2} \\
 + D(\sigma_{x}^{1}\sigma_{y}^{2} - \sigma_{y}^{1}\sigma_{x}^{2})
  + (B+b)\sigma_{z}^{1}    +  (B-b)\sigma_{z}^{2}  ].
\end{multline}

The effective Hamiltonian of four spins is then described as

\begin{equation}
H_{eff} = H \otimes I_4 + I_4 \otimes H,
\end{equation}

\noindent where $I_4$ is the 4 by 4 identity matrix.
The normalized density matrix of thermal entangled system described by $H_{eff}$ in the thermal equilibrium can be found as $\rho_{eff} = e^{-\beta H_{eff}} / tr(e^{-\beta H_{eff}})$, where $tr()$ is the trace function, and $\beta = 1/k T$ with $k$ the Boltzmann constant that we take $k=1$ for simplicity, and $T$ is the temperature.
Note that the second and the third qubits of the thermal entangled system are swapped to obtain the target system for MSG, i.e. $\rho_{eff}\rightarrow SWAP_{2,3}(\rho_{eff})$.
In other words, denoting the system led by $H$ as $\rho$, the target system can be obtained as $\rho_{eff} = SWAP_{2,3}(\rho \otimes \rho)$.
The non-zero elements of $\rho$ are found as \\

\begin{figure}[t!]
  \centering
      \includegraphics[width=0.45\textwidth]{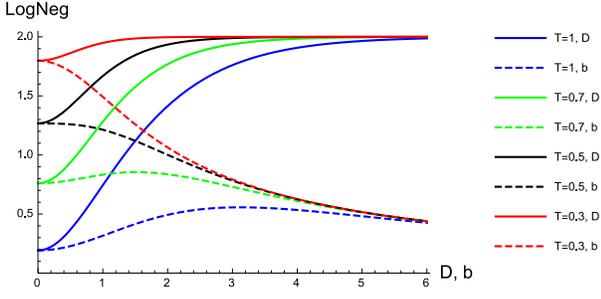}
  \caption{\label{fig:D}(Color online). Entanglement of the thermal entangled state of our system in terms of logarithmic negativity, with respect to Dzyaloshinskii-Moriya (DM) interaction $D$ and external inhomogeneous magnetic field $b$ at various temperatures. Although the amount of entanglement highly depends on the temperature for no or low $D$ or $b$, it becomes independent on the temperature as $D$ or $b$ increases.}
\end{figure}

\noindent
$\rho_{11} =  \gamma^{-1} [\cosh({ B \over k T})  - \sinh({ B \over k T}) ]$,\\
$\rho_{22} =  \gamma      [\cosh({\nu \over k T}) - {b \over \nu} \sinh({\nu \over k T})]$,\\
$\rho_{23} = -\gamma     {1 + i D \over \nu} \sinh({\nu \over k T} )  $,\\
$\rho_{32} = -\gamma     {1 - i D \over \nu} \sinh({\nu \over k T} )  $,\\
$\rho_{33} =  \gamma      [\cosh({\nu \over k T}) + {b \over \nu} \sinh({\nu \over k T})]$, \ \ \\
$\rho_{44} =  \gamma^{-1} [\cosh({ B \over k T})  + \sinh({ B \over k T}) ]$,\\

\noindent where $\gamma = Exp({J_z \over 2 k T})$, $\nu = \sqrt{1 + b^2 + D^2}$ and the trace function yields $2[ \gamma^{-1} \cosh( {\mu \over k T} ) + \gamma \cosh( {\nu \over k T} ) ]$.
It is easy to see that for $D=B=b=0$ and $T$ approaches to zero, $H$ leads to ${1 \over \sqrt{2}}(|01\rangle - |10\rangle)$, and $H_{eff}$ leads to ${1 \over 2}(|0011\rangle +  |1100\rangle - |0110\rangle - |1001\rangle)$, i.e. the success state (after the swap operation).
Now we are ready to analyze the effects of thermalization, DM interaction and magnetic fields, as well as the coupling constant in $z$ direction, on the success probability of MSG.
Note that we take $k=1$ and make the calculations and plot the figures in the units of Boltzmann constant $k$, throughout the paper.

\section{Results and Discussions}
Following the method explained in Section II, we analytically calculate the winning probability $P_w$ for $\rho_{eff}$.
For simplicity, we first choose XX model, i.e. $J_z=0$.
In order to analyze the effect of homogeneous magnetic field $B$ and temperature $T$ on $P_w$, we consider no $D$ and $b$, and find \\

\begin{figure}[t!]
  \centering
      \includegraphics[width=0.45\textwidth]{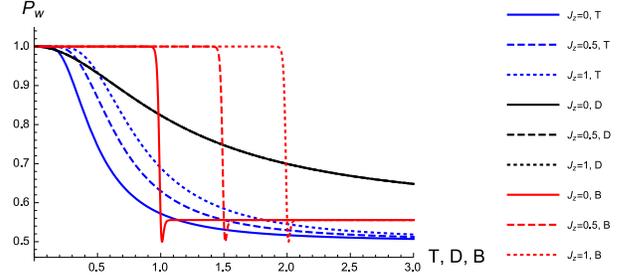}
  \caption{\label{fig:D}(Color online). Probability of winning the magic square game, $P_w$ with respect to temperature $T$ (blue), Dzyaloshinskii-Moriya (DM) interaction $D$ (black) and external homogeneous magnetic field $B$ (red) with $J_z=0$ (solid),  $J_z=0$ (dashed) and  $J_z=0$ (dotted), in the units of Boltzmann constant $k$. Although $P_w$ is not affected by $J_z$ for low temperature and zero $B$, it is easy to check that for high temperature or non-zero $B$, larger $J_z$ makes $P_w$ more robust against $D$ as well.}
\end{figure}

\noindent $ P_w(\rho_{D=b=0}) = {1 \over [\cosh({1 \over T}) + \cosh({B \over T})]^4}  \{ 0.0694 \cosh({4 B \over T}) \\
+ \cosh({3 B \over T}) [0.5 \cosh({1 \over T}) - 0.0555 \sinh({1 \over T})] \\
+ \cosh({2 B \over T}) [-0.0555 \sinh({2 \over T}) + 0.7777 \cosh({2 \over T}) + 0.9444] \\
+\cosh({B \over T})[-0.1111 \sinh({1 \over T})  + 0.0555 \sinh({3 \over T}) \\
+ 2.8888 \cosh({1 \over T})  + 0.6111 \cosh({3 \over T})]  + 0.0277 \sinh({4 \over T}) \\
+ 1.0555 \cosh({2 \over T})  + 0.0972 \cosh({4 \over T}) + 1.05556 \}.$ \\

\noindent As shown in Fig. 1, we find that as temperature decreases, $P_w$ becomes more robust against increasing homogeneous magnetic field $B$, and at the same time it exhibits a sudden change at $B=1$.
For a more clear observation of this sudden change, we focus on the region around $B=1$ in the inset of the Figure.
Note that the this critical point shifts to $B=1.5$ and to $B=2$ for $J_z=0.5$ and for $J_z=1$, respectively, as shown in Fig. 4 (red curves). \\
\indent For a fixed $B$, although $D$ and $b$ lead to different Hamiltonians, density matrices and entanglement values, they yield the same winning probability of the game, i.e.

\noindent $ P_w(\rho_{B=0}) = {sech^2({ \nu \over 2 T}) \over \nu^{3/2} [\cosh({\nu \over T})+1]^2 } [ -0.0833 \nu^2  \sinh({\nu \over T}) \\
+ 0.0277 \nu^2  sinh({3 \nu \over T}) + (0.0694 \nu^2 + 0.0278) \nu \cosh({3 \nu \over T}) \\
+ (0.9305 \nu^2  -0.0278) \nu \cosh({\nu \over T}) \\
+(0.3611 \nu^2  +0.0555) \nu \cosh({2 \nu \over T})  \\
+ (0.6388 \nu^2 - 0.055) \nu ]$,\\

\noindent since $\nu = \sqrt{1 + b^2 + D^2}$.
Therefore for a fixed material, i.e. a fixed strength of DM interaction, the desired effect due to DM interaction can be achieved by applying an external inhomogeneous magnetic field of the same strength, or vice versa, to realize a task with a desired strength of external inhomogeneous magnetic field, it may be possible to choose a material with the same strength of DM interaction.
We plot the dependence of $P_w$ on $D$ (or the same, $b$) for various temperatures in Fig. 2.
Here, although DM interaction excites the entanglement of the system, it decreases the overall success of the task, in contrast to the usual cases that DM interaction both excites the amount of the entanglement of the system and the performance of the task \cite{Zhang2007PRA}.
The physical explanation of this result is that the performance of realizing a quantum task depends not only on the amount of the entanglement but also on the type of it.
In Ref.\cite{Zhang2007PRA} and the following works on the influences of DM interaction, usually systems of two particles were studied and solely the amount of entanglement is sufficient to determine the performance of quantum teleportation or violating Bell's inequality, for instance, leaving room to overlook controversial cases such as the one presented herein.
However, when a system of four particles is considered with a more complex procedure, some of the specific elements of the density matrix of the system become more significant in determining the performance.
That is why an increase in the amount of entanglement of the system alone does not imply an increase in performance of the task.\\
\indent In order to clarify that $D$ and $b$ affects the entanglement of the system oppositely, we plot the logarithmic negativity \cite{LogNeg} of the system with respect to $D$ and $b$ in Fig. 3.
At each temperature value, $D$ increases but $b$ decreases the entanglement of the system and as the strength of $D$ or $b$ increases, dependence of $P_w$ on $T$ decreases.\\
\indent Finally, we analyze the effect of $J_z$ on the winning probability, for three instances, $J_z = 0$, $J_z = 0.5$ and $J_z = 1$ with respect to $T$, $D$ and $B$,  setting other effects to zero for simplicity.
We find that for larger $J_z$, $P_w$ becomes considerably more robust against $B$, more robust against $T$ but it is the same against $D$, as shown in Fig. 4.
However, it is straightforward to see that in the case non zero $B$ or $T$, larger $J_z$ makes $P_w$ more robust against $D$ as well.


\section{Conlusion}
In conclusion, we have studied the influence of DM interaction in a quantum game for the first time, to the best of our knowledge.
We constructed a model of four spins including external homogeneous and inhomogeneous magnetic fields and DM interaction, to serve as the quantum resource to play the magic square game.
Analytically obtaining $P_w$, the probability of winning the game with the thermal entangled state of the spin system, we show how it decreases with respect to increasing temperature and external magnetic fields.
Exciting the entanglement of the systems, although DM interaction excites the performance of realizing tasks in general, here we showed that DM interaction decreases the probability of winning the game.
We also showed that the effects of DM interaction and inhomogeneous magnetic fields on the winning probability are identical.
In addition, we showed that XXZ model is more robust than XX model in playing magic square game.
We believe that our work can be useful in quantum information processing with spin systems.

\section*{Acknowledgements}
This work has been funded by Isik University Scientific Research Funding Agency under Grant Number: BAP 15B103. The author thanks to I. Karakurt for fruitful discussions.


\end{document}